\documentclass{PoS}

\usepackage{enumerate}
\usepackage{amsmath,amssymb}
\usepackage{mathrsfs}
\usepackage{bm}

\title{Physical unitarity of a massive Yang-Mills theory without the Higgs field from a viewpoint of confinement}

\ShortTitle{Physical unitarity of a massive Yang-Mills theory without the Higgs field}

\author{\speaker{Kei-Ichi Kondo}
\\
Department of Physics,  
Graduate School of Science, 
Chiba University, Chiba 263-8522, Japan
\\
        E-mail: \email{kondok@faculty.chiba-u.jp}
}

\author{Kenta Suzuki
\\Department of Physics,  
Graduate School of Science, 
Chiba University, Chiba 263-8522, Japan
\\
}

\author{Hitoshi Fukamachi
\\Department of Physics,  
Graduate School of Science, 
Chiba University, Chiba 263-8522, Japan
\\
}

\author{Shogo Nishino
\\Department of Physics,  
Graduate School of Science, 
Chiba University, Chiba 263-8522, Japan
\\
}

\author{Toru Shinohara
\\Department of Physics,  
Graduate School of Science, 
Chiba University, Chiba 263-8522, Japan
\\
}
        
\abstract{
In this talk, we examine the physical unitarity in a massive Yang-Mills theory without the Higgs field in which the color gauge symmetry is not spontaneously broken and kept intact. 
For this purpose, we use a new framework proposed one of the authors based on a nonperturbative construction of a non-Abelian field describing a massive spin-one vector boson field, which enables us to perform the perturbative and nonperturbative studies on the physical unitarity. 
Moreover, we present a new perturbative treatment for the physical unitarity after giving the general properties of the massive Yang-Mills theory. Then  we reproduce  the violation of physical unitarity in a transparent way.
This work is a preliminary work to the subsequent works in which we present a nonperturbative framework to propose a possible scenario of restoring the physical unitarity in the Curci-Ferrari model. 
We discuss the implications for the low-energy QCD in relation to color confinement, glueball mass and BRST-invariant dimension-two condensate.
}

\FullConference{Xth Quark Confinement and the Hadron Spectrum,\\
		October 8-12, 2012\\
		TUM Campus Garching, Munich, Germany}

\begin{document}

\section{Introduction}

In this talk we reconsider \cite{Kondo12,KSFNS13} a massive Yang-Mills theory \cite{YM54} without the Higgs field \cite{Higgs66}.  
A motivation  of this research stems from  some nonperturbative phenomena caused by strong interactions.
\begin{itemize}
\item[(i)]
Confinement and Green functions--- The deep infrared behaviors of the gluon and ghost Green functions are believed to be intimately connected to color confinement in QCD \cite{KO79,Gribov78}.  
In the Landau gauge, the decoupling solution \cite{decoupling,FMP09,BGP10}  for the gluon and ghost propagators is currently supported rather than the scaling solution \cite{scaling} by recent numerical simulations on large lattices in three and four spacetime dimensions \cite{decoupling-lattice}.
Quite recently, it has been shown \cite{TW11} that the decoupling solution for the gluon and ghost propagators can be well reproduced from a low-energy effective model of a massive Yang-Mills theory, which is a special case of the Curci-Ferrari (CF) model \cite{CF76}.
This feature is not restricted to the Landau gauge and is common to  manifestly Lorentz covariant gauges, e.g., the maximal Abelian gauge \cite{MCM06}, as pointed out and demonstrated in \cite{Kondo11}.
We can ask how color confinement in QCD is understood from the CF model.

\item[(ii)]
Glueball mass spectrum--- 
A glueball should be constructed from the fundamental degrees of freedom of QCD, i.e., quark, gluon and ghost. 
For instance, the potential model of \cite{Cornwall82} identifies   glueballs with bound states of massive gluons.  They are described simply by introducing a naive mass term for gluons, 
$\frac12 M^2 \mathscr{A}_\mu \cdot \mathscr{A}^\mu$, which however breaks the Becchi-Rouet-Stora-Tyutin (BRST) symmetry. 
We ask how we can introduce a BRST-invariant mass term for gluons to establish a firm field theoretical foundation for treating glueballs, which will enable us to answer how precisely the mass and spin of the resulting glueballs are related to those of the constituent gluons. 

\item[(iii)]
Vacuum condensates---  
Besides  gauge-invariant vacuum condensates represented by   $\langle \bar \psi \psi \rangle$ with mass dimension-three and $\langle \mathscr{F}_{\mu\nu}^2 \rangle $ with mass dimension four, which are very important to characterize the nonperturbative vacuum of QCD, there might exist an extra dimension two condensate.  In fact, such a lower dimensional vacuum condensate is needed from the phenomenological point of view.  
However, such a condensate cannot be constructed from  gauge-invariant local composite operators in the framework of the local field theory. 
A  BRST-invariant vacuum condensate of mass dimension two has been  constructed  in  \cite{Kondo01,KMSI02}.  However, it is just  on-shell BRST invariant. 
Can we construct an  off-shell  BRST invariant version of vacuum condensate of mass dimension two?

\end{itemize}

Another motivation of studying the CF model comes from the field theoretical interest, since the massive Yang-Mills theory without the Higgs field has an unsatisfactory aspect   as a quantum field theory.
Renormalizability \cite{tHooft71,tHooft71b} is an important criterion for a quantum field theory to be a calculable and predictable theory. 
In addition, physical unitarity \cite{tHooft71,tHooft71b,KO78,DV70,SF70,Boulware70,CF76b,Ojima82,BSNW96,KG67,DTT88,RRA04} is another important criterion for a quantum theory of gauge fields to be a meaning theory, which prevents unphysical particles from being observed.

In view of this, we remind the readers of the well-known facts:

\begin{itemize}
\item[(i)]
 The massless Yang-Mills theory satisfies both   renormalizability and  physical unitarity \cite{tHooft71,KO78}.

\item[(ii)]
 The massive Yang-Mills theory in which local gauge invariance is spontaneously broken by the Higgs field and the gauge field acquires the mass through the Higgs mechanism satisfies both   renormalizability and   physical unitarity 
\cite{tHooft71b}.
\end{itemize}
In fact, the unified theory of Glashow-Weinberg-Salam for the electromagnetic and weak interactions based on the spontaneous symmetry breaking: $SU(2)_L \times U(1)_Y \rightarrow U(1)_{EM}$ predicted the massive gauge bosons $W^+, W^-$, and $Z^0$ which have been discovered in the mid-1980s, and the remaining Higgs particle is about to be discovered. 

However, in all the models proposed so far as the massive Yang-Mills theory without the Higgs fields (in which the local gauge symmetry is not spontaneously broken), it seems that    renormalizability and   physical unitarity are not compatible with each other.
See \cite{DTT88,RRA04} for reviews and \cite{BFQ} for later developments.
Indeed, the CF model has been shown to be renormalizable \cite{CF76b, BSNW96}, whereas  the CF model does not seem to satisfy   physical unitarity according to \cite{CF76b,Ojima82,BSNW96}.
Although the CF model is not invariant under the usual BRST transformation, it can be made invariant by modifying the  BRST transformation. But, the modified BRST transformation is not nilpotent.

It is known that   nilpotency is the key property to show   physical unitarity in the usual massless Yang-Mills theory,  
since the unphysical states form the BRST quartets and the cancellations occur among the quartets (Kugo-Ojima quartet mechanism) \cite{KO79,KO78}. 
It is not so clear if  nilpotency is necessary to recover physical unitarity in the massive case.
The physical unitarity of the CF model will be discussed in the perturbative and a  nonperturbative framework   in forthcoming papers \cite{Kondo12}.



\section{The Curci-Ferrari model and the modified BRST transformation}

In order to look for a candidate of the massive Yang-Mills theory without the Higgs field, we start from the usual massless Yang-Mills theory in the most general Lorentz gauge formulated in a manifestly Lorentz covariant way.
The total Lagrangian density is written in terms of the Yang-Mills field $\mathscr{A}_\mu$, the  FP ghost field $\mathscr{C}$, the antighost field $\bar{\mathscr{C}}$ and the  NL field $\mathscr{N}$.
As a candidate of the massive Yang-Mills theory without the Higgs field,
we add the ``mass term'' $\mathscr{L}_m$:
\begin{subequations}
	\begin{align}
		\mathscr{L}^{\rm{tot}}_{m\rm{YM}} =& \mathscr{L}_{\rm{YM}} + \mathscr{L}_{\rm{GF+FP}} + \mathscr{L}_{m} , 
\\
		\mathscr{L}_{\rm{YM}}  =& - \frac{1}{4} \mathscr{F}_{\mu \nu} \cdot \mathscr{F}^{\mu \nu} , 
   \\
		\mathscr{L}_{\rm{GF+FP}}  =& \frac{\alpha}{2} \mathscr{N} \cdot \mathscr{N}  + \frac{\beta}{2} \mathscr{N} \cdot \mathscr{N} 
 + \mathscr{N} \cdot \partial^{\mu} \mathscr{A}_{\mu} 
		- \frac{\beta}{2} g \mathscr{N} \cdot (i \bar{\mathscr{C}} \times \mathscr{C}) 
  \nonumber\\
		& + i \bar{\mathscr{C}} \cdot \partial^{\mu} \mathscr{D}_{\mu}[\mathscr{A}] \mathscr{C}
+ \frac{\beta}{4} g^2 (i \bar{\mathscr{C}} \times \mathscr{C}) \cdot (i \bar{\mathscr{C}} \times \mathscr{C}) 
\nonumber\\
		 =& \mathscr{N} \cdot \partial^{\mu} \mathscr{A}_{\mu} + i \bar{\mathscr{C}} \cdot \partial^{\mu} \mathscr{D}_{\mu}[\mathscr{A}] \mathscr{C}
 + \frac{\beta}{4} ( \bar{\mathscr{N}} \cdot \bar{\mathscr{N}} + \mathscr{N} \cdot \mathscr{N}) 
		+ \frac{\alpha}{2} \mathscr{N} \cdot \mathscr{N}
, 
   \\
		\mathscr{L}_{m}  =& \frac{1}{2} M^2 \mathscr{A}_{\mu} \cdot \mathscr{A}^{\mu} + \beta M^2 i \bar{\mathscr{C}} \cdot \mathscr{C} , 
	\end{align}
\end{subequations}
where $\alpha$ and $\beta$ are parameters corresponding to  the gauge-fixing parameters in the $M \rightarrow 0$ limit, 
$
 \mathscr{D}_{\mu}[\mathscr{A}] \mathscr{C}(x) 
:=  \partial_{\mu}\mathscr{C}(x) + g \mathscr{A}(x) \times \mathscr{C}(x)  
$,
 and 
\begin{equation}
\bar{\mathscr{N}} :=-\mathscr{N}+gi\bar{\mathscr{C}} \times \mathscr{C} .
\end{equation}

The $\alpha=0$ case is the  CF  model with the coupling constant $g$, the mass parameter $M$ and the parameter $\beta$.
In the Abelian limit with  vanishing structure constants $f^{ABC}=0$, the FP ghosts decouple and the CF model reduces to the Nakanishi model \cite{Nakanishi72}.

In what follows, we restrict our considerations to the $\alpha=0$ case. 
In the $\alpha=0$ case,   $\mathscr{L}_{\rm YM} + \mathscr{L}_{\rm GF+FP}$ 
is constructed so as to be invariant 
under both the usual BRST transformation and anti-BRST transformation: 
	\begin{align}
		\begin{cases}
			{\boldsymbol \delta}  \mathscr{A}_{\mu}(x) = \mathscr{D}_{\mu}[\mathscr{A}] \mathscr{C}(x)  \\
			{\boldsymbol \delta}  \mathscr{C}(x) = -\frac{g}{2} \mathscr{C}(x) \times \mathscr{C}(x)   \\
			{\boldsymbol \delta}  \bar{\mathscr{C}}(x) = i \mathscr{N}(x)   \\
			{\boldsymbol \delta}  \mathscr{N}(x) = 0  \\
		\end{cases} ,
		\label{BRST}
\quad
\begin{cases}
			\bar{\boldsymbol \delta}  \mathscr{A}_{\mu}(x) = \mathscr{D}_{\mu}[\mathscr{A}] \bar{\mathscr{C}}(x) \\
			\bar{\boldsymbol \delta}  \bar{\mathscr{C}}(x) 
= -\frac{g}{2} \bar{\mathscr{C}}(x) \times \bar{\mathscr{C}}(x)  \\
			\bar{\boldsymbol \delta}   \mathscr{C}(x) = i \bar{\mathscr{N}}(x) \\
			\bar{\boldsymbol \delta}  \bar{\mathscr{N}}(x) = 0
\end{cases} 
 .
\end{align}
Indeed, it is checked that 
\begin{align}
		{\boldsymbol \delta}  \mathscr{L}_{\rm{YM}} = 0 ,
\quad
{\boldsymbol \delta}  \mathscr{L}_{\rm{GF+FP}} = 0 ,
\quad
		\bar{\boldsymbol \delta}  \mathscr{L}_{\rm{YM}} = 0 ,
\quad
\bar{\boldsymbol \delta}  \mathscr{L}_{\rm{GF+FP}} = 0 .
	\end{align}

This is not the case for the mass term $\mathscr{L}_{m}$, i.e., 
	\begin{align}
		{\boldsymbol \delta} \mathscr{L}_{m}  \ne 0 .
	\end{align}
Even in the presence of the mass term $\mathscr{L}_{m}$, however, the  total Lagrangian $\mathscr{L}_{\rm mYM}^{\rm tot}$ can be made invariant by modifying the BRST transformation \cite{CF76}:
$\delta_{\rm BRST}'=\lambda {\boldsymbol \delta}'$ with a Grassmannian number $\lambda$ and
\begin{align}
\begin{cases}
			{\boldsymbol \delta}' \mathscr{A}_{\mu}(x) =  \mathscr{D}_{\mu}[\mathscr{A}] \mathscr{C}(x) \\
			{\boldsymbol \delta}' \mathscr{C}(x) 
=  -\frac{g}{2} \mathscr{C}(x) \times \mathscr{C}(x)  \\
			{\boldsymbol \delta}' \bar{\mathscr{C}}(x) =  i \mathscr{N}(x) \\
			{\boldsymbol \delta}' \mathscr{N}(x) =  M^2 \mathscr{C}(x)  
\end{cases} 
 .
\end{align}
The modified BRST transformation deforms the BRST transformation of the NL field and reduces to the usual BRST transformation in the limit $M \rightarrow 0$.
It should be remarked that 
  ${\boldsymbol \delta}' \mathscr{L}^{\rm tot}_{\rm mYM}=0$ follows from
	\begin{align}
		0 = {\boldsymbol \delta}' (\mathscr{L}_{\rm GF+FP} + \mathscr{L}_{m}) ,
		\label{req}
	\end{align}	
while
 	\begin{equation}
{\boldsymbol \delta}' \mathscr{L}_{m} \ne 0, 
\quad
{\boldsymbol \delta}' \mathscr{L}_{\rm GF+FP} \ne 0.
	\end{equation}

Similarly, the total action is  invariant under a modified anti-BRST transformation $\bar{{\boldsymbol \delta}}'$
defined by 
\begin{align}
\begin{cases}
			\bar{\boldsymbol \delta}' \mathscr{A}_{\mu}(x) = \mathscr{D}_{\mu}[\mathscr{A}] \bar{\mathscr{C}}(x) \\
			\bar{\boldsymbol \delta}' \bar{\mathscr{C}}(x) 
= -\frac{g}{2} \bar{\mathscr{C}}(x) \times \bar{\mathscr{C}}(x)  \\
			\bar{\boldsymbol \delta}'  \mathscr{C}(x) = i \bar{\mathscr{N}}(x) \\
			\bar{\boldsymbol \delta}' \bar{\mathscr{N}}(x) = - M^2 \bar{\mathscr{C}}(x)  
\end{cases} 
 ,
\end{align}
which reduces to the usual anti-BRST transformation in the limit $M \to 0$.
It is sometimes useful to give another form:
\begin{align}
 {\boldsymbol \delta}' \mathscr{\bar N}(x) =  g \mathscr{\bar N}(x) \times \mathscr{C}(x) - M^2 \mathscr{C}(x) , 
\quad
\bar{{\boldsymbol \delta}}' \mathscr{N}(x) =  g \mathscr{N}(x) \times \bar{\mathscr{C}}(x) + M^2 \bar{\mathscr{C}}(x) .
\end{align}

	Moreover, the path-integral integration measure  
$\mathcal{D} \mathscr{A} \mathcal{D} \mathscr{C} \mathcal{D} \bar{\mathscr{C}} \mathcal{D} \mathscr{N}$
is invariant under the modified BRST transformation. 
Indeed, it has been shown in \cite{Kondo12} that the Jacobian associated to the change of integration variables $\Phi(x) \to \Phi'(x) =\Phi(x)+ \lambda {\boldsymbol \delta}' \Phi(x)$ for the integration measure is equal to one. 

However, the modified BRST transformation violates the nilpotency when  $M \not= 0$:
	\begin{align}
\begin{cases}
		{\boldsymbol \delta}' {\boldsymbol \delta}' \mathscr{A}_{\mu}(x) = 0 , \\ 
		{\boldsymbol \delta}' {\boldsymbol \delta}' \mathscr{C}(x) = 0 , \\ 
		{\boldsymbol \delta}' {\boldsymbol \delta}' \bar{\mathscr{C}}(x) = i {\boldsymbol \delta}' \mathscr{N}(x)
		= i M^2 \mathscr{C}(x) \ne 0 , \\ 
		{\boldsymbol \delta}' {\boldsymbol \delta}' \mathscr{N}(x) = M^2 {\boldsymbol \delta}' \mathscr{C}(x)
	=	- M^2 \frac{g}{2} \mathscr{C}(x) \times \mathscr{C}(x) \ne 0 .
\end{cases} 
	\end{align}
The nilpotency is violated also for the modified anti-BRST transformation when  $M \not= 0$:
In the limit $M \to 0$, the modified BRST and anti-BRST transformations reduce  to the usual BRST and anti-BRST transformations and become nilpotent.

\section{Defining a massive Yang-Mills field}

We require the following properties to construct a non-Abelian massive spin-one vector boson field $\mathscr{K}_{\mu}(x)$ in a nonperturbative way:
\renewcommand{\theenumi}{\roman{enumi}}
\renewcommand{\labelenumi}{(\theenumi)}
\begin{enumerate}
\item 
$\mathscr{K}_{\mu}$ has the  modified  BRST invariance (off mass shell): 
	\begin{equation}
		{\boldsymbol \delta}' \mathscr{K}_{\mu} = 0 .
	\end{equation}

\item 
$\mathscr{K}_{\mu}$ is divergenceless (on mass shell): 
	\begin{equation}
		\partial^{\mu} \mathscr{K}_{\mu} = 0 .
	\end{equation}

\item 
$\mathscr{K}_{\mu}$ obeys the adjoint transformation under the color rotation:
	\begin{equation}
		\mathscr{K}_{\mu}(x) \to U \mathscr{K}_{\mu}(x) U^{-1} , \quad U = \exp[i \varepsilon^A Q^A] ,
	\end{equation}	
\end{enumerate}
	which has the infinitesimal version: 
	\begin{equation}
		\delta \mathscr{K}_{\mu}(x) = \varepsilon \times \mathscr{K}_{\mu}(x) . 
	\end{equation}
The field $\mathscr{K}_\mu$ is identified with the non-Abelian version of the physical massive vector field with spin one, as ensured by the above properties.
Here (i) guarantees that $\mathscr{K}_{\mu}$ belong  to the physical field creating a physical state with positive norm.
(ii)  guarantees that $\mathscr{K}_{\mu}$ have the correct degrees of freedom as a massive spin-one particle, i.e., three in the four-dimensional spacetime, i.e., two transverse and one longitudinal modes, excluding one scalar mode. 
(iii) guarantees that $\mathscr{K}_{\mu}$ obey the same transformation rule as that of the original gauge field $\mathscr{A}_{\mu}$

We observe that the total Lagrangian of the CF model is invariant under the (infinitesimal) \textbf{global gauge transformation} or \textbf{color rotation} defined by
	\begin{align}
		&\delta \Phi(x) := [\varepsilon^C i Q^C, \Phi(x)] = \varepsilon \times \Phi(x) , 
 {\rm for} \quad
		 \Phi=\mathscr{A}_{\mu}, \mathscr{N}, \mathscr{C} , \bar{\mathscr{C}} ,  \\
		&\delta \varphi(x) := [\varepsilon^C i Q^C, \varphi(x)] = - i \varepsilon \varphi(x) ,
	\end{align}
where $\varphi$ is a matter field.
The  conserved Noether charge $Q^A := \int d^3x \mathscr{J}^{0,A}_{\rm color}$ obtained from the color current $\mathscr{J}^0_{\rm color}$ is called the \textbf{color charge}
and is equal to the generator of the color rotation.

It has been shown \cite{Kondo12} that such a field $\mathscr{K}_{\mu}$ is obtained by a nonlinear but local transformation from the original fields  $\mathscr{A}_\mu$,  $\mathscr{C}$,  $\bar{\mathscr{C}}$ and $\mathscr{N}$ of the CF model:
\begin{align}
 \mathscr{K}_\mu :=&  \mathscr{A}_\mu - M^{-2} \partial_\mu \mathscr{N} 
- gM^{-2} \mathscr{A}_\mu \times \mathscr{N} 
\nonumber\\
&+ gM^{-2}  \partial_\mu \mathscr{C} \times i\bar{\mathscr{C}} 
+ g^2 M^{-2} (\mathscr{A}_\mu \times \mathscr{C}) \times i \bar{\mathscr{C}} .
\label{K}
\end{align}
In the Abelian limit or the lowest order of   $g$, $\mathscr{K}_{\mu}$ reduces to the Proca field for massive vector:
	\begin{equation}
		\mathscr{K}_{\mu} \to \mathscr{A}_{\mu} - \frac{1}{M^2} \partial_{\mu} \mathscr{N} := U_{\mu} .
	\end{equation}
It should be remarked that $U_\mu$	is invariant under the Abelian version of the modified BRST, but it is not invariant under the non-Abelian modified BRST transformation.

The new field $\mathscr{K}_{\mu}$ is converted to a simple form:
	\begin{equation}
		\mathscr{K}_{\mu}(x) = \mathscr{A}_{\mu}(x) + \frac{1}{M^2} i {\boldsymbol \delta}' \bar{{\boldsymbol \delta}}' \mathscr{A}_{\mu}(x) .
		\label{K2}
	\end{equation}
It has been explicitly shown in \cite{Kondo12} that the field $\mathscr{K}_{\mu}$ defined by (\ref{K}) or (\ref{K2}) satisfies all the above properties.
The field $\mathscr{K}_{\mu}$ plays the role of the non-Abelian massive vector field
and is identified with a non-Abelian version of the spin-one massive vector field.
Equation (\ref{K}) gives a transformation from $\mathscr{A}_{\mu}, \mathscr{N}, \mathscr{C}$ and $\bar{\mathscr{C}}$ to $\mathscr{K}_{\mu}$.

As an  application of the above result, we can construct a mass term which is invariant simultaneously under the modified BRST transformation,  Lorentz transformation and  color rotation:
	\begin{equation}
		\frac{1}{2} M^2 \mathscr{K}_{\mu}(x) \cdot \mathscr{K}^{\mu}(x) .
	\end{equation}
This can be useful as a regularization scheme for avoiding infrared divergences in non-Abelian gauge theories. 
Moreover, we can obtain a dimension-two condensate which is modified  BRST invariant, Lorentz invariant, and color-singlet: 
	\begin{equation}
		\langle \mathscr{K}_{\mu}(x) \cdot \mathscr{K}^{\mu}(x) \rangle .
	\end{equation}
	This dimension-two condensate is off-shell (modified) BRST invariant and  should be compared with the dimension-two condensate proposed in \cite{Kondo01,KMSI02} which is only on-shell BRST invariant:
	\begin{equation}
		\Big\langle \frac12 \mathscr{A}_{\mu}(x) \cdot \mathscr{A}^{\mu}(x) + \beta  \mathscr{C}(x) \cdot \mathscr{\bar C}(x) \Big\rangle .
	\end{equation}

\section{Perturbative violation of physical unitarity}

In \cite{KSFNS13} , we  have checked in a new perturbative treatment whether or not the CF model satisfies the physical unitarity.
Then we have confirmed the violation of the physical unitarity in the perturbative treatment and we have clarified the reason  in the massive Yang-Mills theory without the Higgs field. 
The perturbative analysis for the   physical unitarity imposes a restriction on the valid energy together with  the parameter of the CF model:
$E^2 < 4\beta M^2$ 
in order to confine unphysical modes (ghost, antighost, scalar mode).
However, $\beta=0$ is not allowed in this scenario.

It should be remarked that even the modified BRST (and anti-BRST) invariant quantity depends on a parameter $\beta$ in the $M \not= 0$ case.  
This should be compared with the  $M=0$  case, in which $\beta$ is a gauge-fixing parameter and the BRST-invariant quantity does not depend on $\beta$, which means that the physics does not depend on $\beta$ in the $M=0$ case. 
This is not the case for $M \not= 0$  \cite{Lavrov12}.



\section{Possible nonperturbative restoration of physical unitarity}

The conclusion  obtained in this work still leaves  a possibility that the nonperturbative approach can modify the conclusion.
In a subsequent paper, indeed, we will propose  a scenario in which the physical unitarity can be recovered in the CF model thanks to the FP conjugation invariance. 
Indeed, we will show that the norm cancellation is automatically guaranteed from the Slavnov-Taylor identities if the ghost-antighost bound state exists. 
In this way, the physical unitarity can be recovered in a nonperturbative way. 
To show the existence of the bound state of ghost and antighost,  the Nambu-Bethe-Salpeter equation is to be solved. This is a hard work to be tackled in subsequent papers.

{\it Acknowledgements}:\ 
This work is  supported by Grant-in-Aid for Scientific Research (C)  24540252 from the Japan Society for the Promotion of Science (JSPS).


\begin{thebibliography}{99}
\bibitem{Kondo12} 
K.-I. Kondo,
arXiv:1208.3521[hep-th],
Phys. Rev. D\textbf{87}, 025008 (2013).


\bibitem{KSFNS13} 
K.-I. Kondo, K. Suzuki, H. Fukamachi, S.Nishino, and T. Shinohara, 
arXiv:1209.3994 [hep-th],
Phys. Rev. D\textbf{87}, 025017 (2013).




\bibitem{YM54}
C.N. Yang and R.L. Mills,
Phys. Rev. {\bf 96}, 191
 (1954).


\bibitem{Higgs66}
P.W. Higgs,
Phys. Rev. {\bf 145}, 1156
 (1966).


\bibitem{KO79}
 T. Kugo and I. Ojima,
Suppl. Prog. Theor. Phys. \textbf{66}, 1
(1979).


\bibitem{Gribov78}
V.N. Gribov. 
Nucl Phys. B\textbf{139}, 1
 (1978).


\bibitem{decoupling}
 Ph. Boucaud, J.P. Leroy, A. Le Yaouanc, J. Micheli, O. Pene and J. Rodriguez-Quintero, 
[hep-ph/0803.2161],
JHEP 0806, 099 (2008).
\\
 A.C. Aguilar, D. Binosi and J. Papavassiliou,
arXiv:0802.1870[hep-ph],
Phys.Rev.D{\bf 78}, 025010 (2008).


\bibitem{FMP09}
C.S.~Fischer, A.~Maas and J.M.~Pawlowski,
arXiv:0810.1987 [hep-ph],
Annals Phys.\textbf{324}, 2408
 (2009).


\bibitem{BGP10}
J. Braun, H. Gies and J.M. Pawlowski, 
arXiv:0708.2413 [hep-th], 
Phys. Lett. B{\bf 684}, 262
 (2010). 


\bibitem{scaling}
R. Alkofer and L. von Smekal, 
[hep-ph/0007355], 
Phys. Rep. \textbf{353}, 281
 (2001).


\bibitem{decoupling-lattice}
I.L. Bogolubsky, E.M. Ilgenfritz, M. Muller-Preussker and A. Sternbeck, 
arXiv:0710.1968 [hep-lat], 
PoS LAT2007, 290 (2007). 
\\
A. Cucchieri and T. Mendes,
arXiv:0710.0412 [hep-lat], 
PoS LAT2007, 297 (2007). 
\\
A. Sternbeck, L. von Smekal, D.B. Leinweber  and A.G. Williams, 
arXiv:0710.1982 [hep-lat] 
PoS LAT2007, 340 (2007). 



\bibitem{TW11}
M. Tissier and N. Wschebor,  
arXiv:1105.2475 [hep-th], 
Phys. Rev. D\textbf{84}, 045018 (2011). 
\\
M. Tissier and N. Wschebor,  
arXiv:1004.1607 [hep-ph], 
Phys. Rev. D\textbf{82}, 101701 (2010). 
\\
J. Serreau and M. Tissier,
arXiv:1202.3432 [hep-th], 
Phys.Lett. B\textbf{712},  97--103 (2012).


\bibitem{CF76}
G. Curci and R. Ferrari,
Nuovo Cim. A{\bf 32}, 151
 (1976).


\bibitem{MCM06}
T. Mendes, A. Cucchieri and A. Mihara, 
[hep-lat/0611002], 
AIP Conf. Proc.{\bf 892}, 203
 (2007).


\bibitem{Kondo11}
K.-I. Kondo,
arXiv:1103.3829 [hep-th], 
Phys. Rev. D\textbf{84}, 061702 (2011). 


\bibitem{Cornwall82}
J.M. Cornwall and A. Soni,  
Phys. Lett. B{\bf 120}, 431
 (1983). 


\bibitem{Kondo01}
K.-I. Kondo,
[hep-th/0105299], 
Phys. Lett. B\textbf{514}, 335
 (2001). 
\\
K.-I. Kondo,
[hep-th/0306195], 
Phys. Lett. B\textbf{572}, 210
 (2003). 


\bibitem{KMSI02}
K.-I. Kondo, T. Murakami, T. Shinohara and T. Imai, 
[hep-th/0111256], 
Phys. Rev. D\textbf{65}, 085034 (2002). 


\bibitem{tHooft71}
G. 't Hooft,
Nucl. Phys. B{\bf 33}, 173
 (1971).


\bibitem{tHooft71b}
G. 't Hooft,
Nucl. Phys. B{\bf 35}, 167
 (1971).


\bibitem{KO78}
T. Kugo and I. Ojima,
Phys. Lett. B{\bf 73}, 459
 (1978).


\bibitem{DV70}
H. van Dam and M.J.G. Veltman,
Nucl. Phys. B{\bf 22}, 397
 (1970). 


\bibitem{SF70}
A.A. Slavnov and L.D. Faddeev, 
Theor. Math. Phys. {\bf 3},  312
 (1970) [Teor. Mat. Fiz. {\bf 3}, 18
 (1970)]. 
\\
A.A. Slavnov,
Theor. Math. Phys. {\bf 10},  201
 (1972) [Teor. Mat. Fiz. {\bf 10}, 305
 (1972)]. 


\bibitem{Boulware70}
D.G. Boulware, 
Annals Phys. {\bf 56}, 140
 (1970). 


\bibitem{CF76b}
G. Curci and R. Ferrari, 
Nuovo Cim. A{\bf 35}, 1
 (1976), Erratum-ibid. A{\bf 47}, 555 (1978). 


\bibitem{Ojima82}
I. Ojima, 
Z. Phys. C{\bf 13}, 173
 (1982).


\bibitem{BSNW96}
J. de Boer, K. Skenderis, P. van Nieuwenhuizen and A. Waldron,
Phys. Lett. B{\bf 367},  175
 (1996).
 

\bibitem{KG67}
T. Kunimasa and T. Goto,  
Prog. Theor. Phys. {\bf 37}, 452
 (1967). 
\\
T. Fukuda, M. Monda, M. Takeda and Kan-ichi Yokoyama,
Prog. Theor. Phys. {\bf 66}, 1827
 (1981).
\\
J.M. Cornwall, 
Phys. Rev. D{\bf 26}, 1453
 (1982). 


\bibitem{DTT88}
R. Delbourgo, S. Twisk and G. Thompson,
Int. J. Mod. Phys. A{\bf 3}, 435
 (1988).


\bibitem{RRA04}
H. Ruegg and M. Ruiz-Altaba,
[hep-th/0304245],
Int. J. Mod. Phys. A{\bf 19},  3265
 (2004).


\bibitem{BFQ}
R. Ferrari,   
arXiv:1106.5537 [hep-ph],
Acta Phys.Polon. B\textbf{43}, 1735-1767  (2012).







\bibitem{Nakanishi72}
N. Nakanishi,
Phys. Rev. D{\bf 5},  1324
 (1972).


\bibitem{Lavrov12}
P.M. Lavrov,
arXiv:1205.0620 [hep-th], 
Mod.Phys.Lett. A\textbf{27}, 1250132 (2012). 



 
\end{thebibliography}
\end{document}